# Phoneme-Based Ratio Mask Estimation for Reverberant Speech Enhancement in Cochlear Implant Processors

Kevin M. Chu, Leslie M. Collins, and Boyla O. Mainsah

Department of Electrical and Computer Engineering, Duke University, Durham, NC, USA

*Abstract*— **Cochlear implant (CI) users have considerable difficulty in understanding speech in reverberant listening environments. Time-frequency (T-F) masking is a common technique that aims to improve speech intelligibility by multiplying reverberant speech by a matrix of gain values to suppress T-F bins dominated by reverberation. Recently proposed mask estimation algorithms leverage machine learning approaches to distinguish between target speech and reverberant reflections. However, the spectro-temporal structure of speech is highly variable and dependent on the underlying phoneme. One way to potentially overcome this variability is to leverage explicit knowledge of phonemic information during mask estimation. This study proposes a phoneme-based mask estimation algorithm, where separate mask estimation models are trained for each phoneme. Sentence recognition tests were conducted in normal hearing listeners to determine whether a phoneme-based mask estimation algorithm is beneficial in the ideal scenario where perfect knowledge of the phoneme is available. The results showed that the phoneme-based masks improved the intelligibility of vocoded speech when compared to conventional phoneme-independent masks. The results suggest that a phoneme-based speech enhancement strategy may potentially benefit CI users in reverberant listening environments.**

## I. INTRODUCTION

COCHLEAR implants (CIs) aim to restore functional levels of speech perception and speech recognition to individuals with profound sensorineural hearing loss [1]. A CI processor encodes the spectro-temporal content of a speech signal as a discrete sequence of electrical pulses that stimulate frequency-specific regions of the cochlea. While CI users typically have high speech recognition in anechoic listening environments [1], they have considerably more difficulty in understanding speech in listening environments that contain reverberation and noise [2], [3]. Reverberation refers to the sum total of delayed and attenuated copies of the direct signal that result from reflections from walls and other objects in an enclosed space [4].

Reverberation flattens formant transitions [5], fills silent gaps with reverberant reflections [2], masks low-energy phonemes [2], and reduces temporal envelope modulations, which are all important for speech intelligibility [6]. Noise also masks low-energy phonemes and distorts the temporal envelope, but unlike reverberation these effects do not depend on the energy of the preceding speech segments [6]. Reverberation and noise are more detrimental to a CI user than to a normal hearing listener because a CI user's spectral resolution is limited by the number of electrodes in the array [2] and by interactions between electrode channels [7].

A common speech enhancement algorithm is time-frequency (T-F) masking, where the T-F representation of corrupted speech is multiplied by a matrix of gain values to attenuate T-F bins dominated by reverberation and noise [8]. One type of T-F mask is the binary mask, where the mask is assigned a value of 1 if the signal in a T-F bin satisfies some criterion, or 0 if the criterion is not satisfied [9]. A common criterion is whether the signal-to-noise ratio (SNR) or signal-to-reverberant ratio (SRR) exceeds a predefined threshold, such that T-F bins with high amounts of distortion are removed while speech-dominated T-F bins are retained [2], [9]. Another type of T-F mask is the ratio mask, where the gain values range continuously between 0 and 1 [10], [11]. In contrast to the binary mask, the ratio mask aims to attenuate rather than completely remove highly distorted T-F bins [10].

In a real-time scenario where the mask is unavailable [2], an algorithm is required to estimate the mask based only on features extracted from the reverberant signal. Traditional algorithms attempt to estimate the mask using statistical signal processing techniques. One such algorithm estimates the binary mask by comparing a kurtosis-based feature to an adaptive threshold [12], as kurtosis is generally smaller for reverberant speech than anechoic speech [13]. However, this algorithm cannot be implemented in real-time because it requires non-causal acoustic information to calculate the kurtosis-based feature [12]. A second algorithm estimates the binary mask based on the ratio between the power spectrum of the linear prediction residual to that of the reverberant signal, as low

The authors are with the Department of Electrical and Computer Engineering, Duke University, Durham, NC 27708 USA (e-mail: kevin.m.chu@duke.edu, leslie.collins@duke.edu, boyla.mainsah@duke.edu).

values of this ratio indicate the presence of formants [14]. However, this algorithm requires model parameters to be empirically tuned for each test environment, which is unknown in a real-time scenario [14]. A third algorithm estimates the ratio mask based on recursive estimates of the signal-to-distortion ratio, which identifies speech-dominated regions of the reverberant signal [15]. However, this algorithm calculates the signal-to-distortion ratio in a manner that assumes the existence of an initial segment of the signal that contains noise but not reverberation, but this initial segment is not guaranteed in a real-time setting [15]. Given the limitations of these statistical signal processing-based algorithms, an alternative approach is preferred.

Recently proposed mask estimation algorithms leverage machine learning to automatically recognize patterns in the reverberant signal than can be used to distinguish target speech from reverberation and noise. Machine learning algorithms such as deep neural networks (DNNs) and recurrent neural networks (RNNs) have been shown to improve speech intelligibility in non-stationary noises in both hearing aid users [16], [17] and CI users [18]. In addition, machine learning algorithms have improved speech intelligibility in hearing aid users in listening environments that contain both reverberation and noise [19]–[21]. However, these algorithms were trained and tested on the same speaker [19], [20] or same reverberant room [19]–[21], so their reported benefits were potentially over-estimated.

To the best of our knowledge, only one machine learning-based mask estimation algorithm has been proposed for CI users in reverberant listening environments. Previous work by Desmond used relevance vector machines (RVMs) to detect and mitigate late reverberant reflections within each electrode channel of a CI [22], as mitigating the late reflections leads to substantial improvements in speech intelligibility in CI users [23]–[25]. As classification is imperfect, Desmond attempted to achieve a balance between detections and false alarms by setting a classifier operating point that remained fixed across electrodes [26]. However, this algorithm did not consistently improve speech intelligibility in reverberant environments that were not used in the training data [26]. Desmond hypothesized that the algorithm could potentially improve intelligibility if the classifier operating point were allowed to vary across electrodes to better account for the frequency-dependent effect of reverberation [26]. Fig. 1 shows an electrodogram of the HINT sentence "the boy broke the wooden fence" [27] in a reverberant listening environment. In high frequencies, reverberant reflections decay exponentially over time (e.g. the instance of /s/ after 1500 ms in Fig. 1), so they can be distinguished from speech. In low frequencies, reverberant reflections are often interrupted by the following phoneme before they can fully decay (e.g. portions of the electrodogram prior to 1500 ms in Fig. 1), so it is much more difficult to differentiate between reverberant reflections and target speech [26].

In addition to frequency dependence, the effects of reverberation also depend on the underlying phoneme. Vowels are primarily affected by self-masking [28], which occurs when reverberant reflections of a phoneme smear the internal energy within that phoneme [29]. Self-masking in vowels flattens formant transitions, which causes diphthongs to be confused as monophthongs [5]. Consonants are primarily affected by overlap-masking [28], where the reverberant reflections of a phoneme overlap with active speech segments of the following phoneme [29]. Overlap-masking causes low-energy consonants to be masked by high-energy reverberant reflections, which results in the loss of consonant information [28]. Overlap-masking is especially detrimental to stop consonants as the reverberant reflections fill in stop closures, which are an important cue for perceiving stop consonants [6], [28].

Given that reverberation exerts different effects on different phonemes, one potential way to improve mask estimation is to leverage explicit knowledge of the underlying phoneme. The current study proposes a phoneme-based mask estimation algorithm, where a separate RNN-based mask estimation model is trained for each phoneme. As shown in Fig. 2, phonemes are typically concentrated in specific frequency ranges, with vowels containing primarily low frequency content and fricatives containing high frequencies [30]. Thus, each phoneme-specific model operates on a smaller acoustic feature

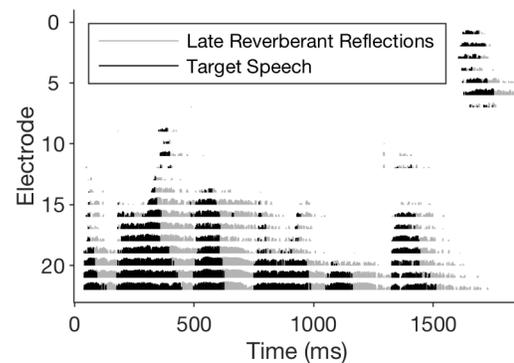

Fig. 1. Electrodogram of the HINT sentence "the boy broke the wooden fence" in a reverberant environment. An electrodogram is a visualization of the electrical pulses delivered by each electrode channel as a function of time. The horizontal axis is time and vertical axis is electrode (frequency increases as the electrode index decreases). The tick marks represent electrical pulses that were present in both the reverberant signal and the corresponding anechoic signal, and the gray tick marks denote pulses that represent instances where silent gaps were filled in with reverberation reflections.

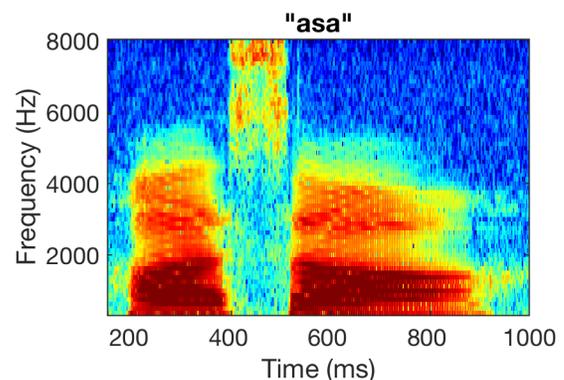

Fig. 2. Spectrogram of the speech token "asa". As the figure illustrates, different phonemes have different spectra. For instance, /a/ contains primarily low frequency content, and /s/ contains primarily high frequency content.

TABLE I
RIR CHARACTERISTICS

| Dataset | Room | Dimensions (L x W x H) (m) | Source-Receiver Distance (m) | $RT_{60}$ (s) | DRR (dB) |
|---|---|---|---|---|---|
| Training & Validation (Simulated) | Meeting | 3.6 x 4.4 x 2.7 | 1.0 | 0.3 | -1.1 |
| | | | 3.0 | 0.3 | -3.4 |
| | Seminar | 8.6 x 7.8 x 2.7 | 1.0 | 0.5 | 3.8 |
| | | | 3.0 | 0.5 | -3.5 |
| | Auditorium | 15.8 x 11.7 x 7.4 | 1.0 | 1.7 | 3.5 |
| | | | 3.0 | 1.7 | -4.2 |
| | | | 6.0 | 1.7 | -8.4 |
| | Lecture | 7.4 x 7.4 x 3.0 | 1.3 | 0.5 | 0.1 |
| | | | 2.6 | 0.5 | -4.0 |
| | | | 5.2 | 0.5 | -7.6 |
| | Kitchen | 7.4 x 7.4 x 3.0 | 1.3 | 0.7 | -2.6 |
| | | | 2.6 | 0.7 | -6.5 |
| | | | 5.2 | 0.7 | -9.2 |
| | Office | 12.2 x 12.2 x 3.0 | 1.3 | 1.0 | 0.4 |
| | | | 2.6 | 1.0 | -3.6 |
| | | | 5.2 | 1.0 | -5.8 |
| Testing (Recorded) | Office | 5.0 x 6.4 x 2.9 | 3.0 | 0.6 (left) | 0.2 (left) |
| | | | | 0.6 (right) | 0.4 (right) |
| | Stairway | 7.0 x 5.2 | 3.0 | 1.0 (left) | 1.9 (left) |
| | | | | 0.9 (right) | 1.6 (right) |
| | Aula Carolina | 19.0 x 30.0 | 5.0 | 6.6 (left) | -1.5 (left) |
| | | | | 6.5 (right) | -0.6 (right) |

This table describes the RIR type (simulated or recorded), the room type, the dimensions, the source-receiver distance, the reverberation time, and the direct-to-reverberant ratio. The recorded RIRs used in the testing set were binaural, recorded from right and left channels of a dummy head.

space as compared to a phoneme-independent model. This algorithm is similar to the Mixture of Experts approach, where the input feature space is divided into different regions containing a different model that is an expert at solving the task within that region [31]. A phoneme-specific mask estimation algorithm was shown to improve the performance of an automatic speech recognition (ASR) model in a noisy and reverberant environment as compared to a phoneme-independent algorithm [32]. However, this algorithm was designed to remove only the additive noise, so it is unclear whether a mask estimation algorithm that aims to remove reverberation can also benefit from a phoneme-based approach. Suppressing reverberation is especially difficult because reverberant reflections are correlated with anechoic speech [12], so the current study focuses on utilizing phoneme-based masks to remove reverberation. Furthermore, it is not known whether a phoneme-based approach to T-F masking is beneficial for CI users.

To determine whether phoneme-specific T-F mask estimation is beneficial for CI users, the first step is to obtain an upper bound by assuming ideal knowledge of the phoneme. We hypothesize that phoneme-based mask estimation with ideal phonemic knowledge will provide greater benefits in speech intelligibility as compared to a phoneme-independent model. The phoneme-based framework was evaluated by conducting speech intelligibility tests in normal hearing listeners using vocoded speech.

## II. METHODS

### A. Speech Stimuli

All mask estimation models were trained using utterances from the LibriSpeech corpus of spoken English [33]. Speech enhancement algorithms trained on LibriSpeech have been shown to generalize across different speech corpora due to its large number of speakers and large variation in recording channel [34]. The training set for the mask estimation algorithms was generated by randomly selecting 3696 utterances from the 100-hour LibriSpeech training set with a roughly equal number of utterances selected for each of the 251 speakers. The resulting training set contained a total of 13.1 hours of data recorded at a sampling frequency of 16 kHz. The anechoic utterances were convolved with simulated room impulse responses (RIRs) that were generated via the Modified Image Source Method [35], [36]. The Modified Image Source Method models reverberant reflections as emanating from image sources placed within the exposed surfaces of an acoustic environment [35], [36]. The simulated RIRs were generated using a sampling frequency of 48 kHz, filtered using an anti-aliasing filter, and then downsampled to 16 kHz prior to convolution with the anechoic speech stimuli. The RIRs were simulated based on the meeting room, seminar room, and auditorium from [25], as well as the lecture room, kitchen, and office from [37]. For each room, the source was placed at the front center of room 1m away from the wall, and the height was randomly selected from a uniform distribution in the range 1m to 2m to simulate the height of a human speaker. The receiver was also placed at the center of the room at varying source-receiver distances and at the same height as the source. Table I lists the RIR characteristics, including room dimensions, source-receiver distances, reverberation times ($RT_{60}$), and direct-to-reverberant ratios (DRR). The room dimensions, source-receiver distances, and $RT_{60}$s were taken from [25], [37]. For the lecture room, kitchen, and office from [37], only the room volume was provided, but the room dimensions were not. A height of 3m was arbitrarily assumed, as this is similar to the heights of the meeting room and seminar room from [25]. The length and width of each room were assumed to be equal and were calculated based on the provided room volume and the assumed height. The DRRs were calculated by computing the ratio between the total energy of the *direct path* component of the RIR to the energy of the reverberant reflections [4]. The direct path signal refers to the component of the reverberant signal that travels directly from the source to the receiver without reflecting off the surfaces in the room [4]. The direct path component was defined as the initial portion of the RIR up through 8ms following the arrival time of the direct sound [4], which was estimated by dividing the source-receiver distance by the speed of sound (assumed to be 343m/s). The development set was generated by randomly selecting 400 utterances from the LibriSpeech dev-clean set (~0.9 hours) and convolving them with the same RIRs used for the training data.

The models were tested using sentences from the Hearing in Noise Test (HINT) database [27]. Three test sets were created by convolving anechoic sentences with recorded RIRs from three rooms in the Aachen Impulse Response database [38]. RIRs were selected from an office, stairway, and Aula Carolina [38], [39]. For the stairway and Aula Carolina, RIRs were selected at an azimuth of 90 degrees, where the source and receiver are directly facing each other. To match the sampling frequency of the HINT sentences, the RIRs were filtered using an anti-aliasing filter and then downsampled from 48 kHz to 16 kHz prior to convolution with the anechoic speech stimuli. Table I shows the characteristics of the RIRs in the testing dataset. The $RT_{60}$s were calculated based on the Schroeder method [40] using code provided by [41]. The DRRs for the recorded RIRs were calculated using a slightly different definition for the direct path, which was defined as 8ms following the largest amplitude filter coefficient [4].

### B. Conventional Mask Estimation

The target is the ideal ratio mask (IRM), where the gain values range continuously from 0 to 1 [11]. The IRM was originally developed to remove additive noise, and is computed as shown in (1).

$$IRM(t,f) = \sqrt{\frac{|X(t,f)|^2}{|X(t,f)|^2 + |N(t,f)|^2}} \quad (1)$$

where $|X(t,f)|^2$ represents the power spectrum of clean speech at time $t$ and frequency bin $f$; and $|N(t,f)|^2$ represents the power spectrum of the noise. The enhanced signal is calculated via pointwise multiplication between the ratio mask and the magnitude spectrum of noisy speech [42]. The IRM can be adapted for use in reverberant environments by defining the clean signal as the direct path signal and the noise as the reverberant residual. The direct path signal was generated by convolving the anechoic signal with the direct path component of the RIR [4]. The reverberant residual refers to the difference between the reverberant signal and the direct path signal. The ratio mask was used instead of the binary mask because the ratio mask does not require an SRR-based threshold to be tuned [18].

Both the acoustic features and IRM were calculated at the time-frequency resolution of a CI processor to avoid having to resynthesize the time domain signal using the reverberant phase. The features and IRM were calculated according to the Advanced Combination Encoder (ACE) strategy [43] based on a Nucleus CI with 22 electrodes. Acoustic features were extracted by buffering the time domain signal into 8ms frames with a 2ms shift between consecutive frames. The buffered signal was then passed through a short-time Fourier transform (STFT), resulting in a set of 65 frequency bins from which a power spectrum was computed. The power spectrum was logarithmically compressed to reduce the dynamic range, and the log compressed spectrum was used as the features for the mask estimation models. The features were then normalized based on the mean and variance of the training dataset. The IRM was calculated from the uncompressed 65-dimensional STFT power spectra of the reverberant signal and corresponding direct path signal. Both the acoustic features and IRM were calculated at an early stage of signal processing prior to applying the 22 weighted ACE filters and the user-specific map, so this approach is potentially applicable to CIs that use different processing strategies.

The mask estimation models were trained using the PyTorch library [44]. As baseline phoneme-independent mask estimation models, we used neural networks with one and two long short-term memory (LSTM) [45] layers. These baselines were based on the model proposed by [18], which was developed specifically to estimate the ratio mask in non-stationary noise for CI applications. Each LSTM layer consisted of 128 memory blocks containing input, forget, and output gates with sigmoid activations as well as a cell state. The memory blocks did not use peephole connections or projection layers. The last LSTM layer was fully connected to an output layer with 65 sigmoidal units, with each unit representing the value of the mask at each STFT bin. The network was trained over a causal contextual window with a width of five frames [18]. Because the objective is a real-valued mask, the network was trained to minimize the mean squared error (MSE) between the predicted and the ideal mask. The weights and biases of both the LSTM layers and the fully connected layers were randomly initialized from a uniform distribution in the range [-0.1, 0.1]. The network was trained using the Adam optimizer [46], with a batch size of 2 utterances, a learning rate of 1e-3, and exponential decay rates of 0.9 and 0.999 for the first and second moments, respectively. The network was trained until the MSE loss on the development set did not decrease by more than 0.001 over the last 10 epochs, at which point training was halted. The models were tested on reverberant sentences from the HINT database [27], as described previously. All experiments were carried out using an NVIDIA Titan V graphics processing unit (GPU).

*C. Phoneme-Dependent Mask Estimation*

The phoneme-specific models were initialized using the weights from the trained phoneme-independent model with one LSTM layer. Each phoneme-specific model was trained further using only time frames corresponding to the relevant phoneme. As with the phoneme-independent models, each phoneme-specific model was trained until the MSE failed to decrease by more than 0.001 over the last 10 epochs. The phoneme labels were automatically generated using forced alignment, which determines the time stamps where phonemes occur within an acoustic signal. Forced alignments were generated with a Gaussian Mixture Model-Hidden Markov Model-based aligner using the LibriSpeech recipe [33] from the Kaldi ASR toolkit [47]. The forced aligner was trained using 13-dimensional mel-frequency cepstral coefficients (MFCCs), deltas, and delta-deltas extracted over 25ms frames with a 10ms frame shift. The features were mean and variance normalized on a per-speaker basis, spliced across +/-3 time frames, reduced to 40 dimensions using linear discriminant analysis, decorrelated using maximum likelihood linear transformation, and speaker adapted using feature-space maximum likelihood linear regression. To obtain high quality phoneme labels, forced alignments were generated from the direct path component of the reverberant utterances used in the mask estimation training set. To ensure compatibility with the mask estimation models, the phoneme time stamps obtained from the forced aligner were converted to CI-based time bins. The models were tested on sentences in the HINT database, using forced alignment to generate ideal phoneme labels.

In addition to phoneme-specific masks, this study considers masks based on groups of phonemes. While phoneme-specific models can overcome the large spectro-temporal variability in speech, there are some potential limitations. The first limitation is that phoneme-specific masks require the training data to be split into phoneme-specific subsets, which greatly reduces the amount of data on which each model is trained. Another limitation is that phoneme classification is a relatively challenging task because phonemes with similar spectro-temporal characteristics are often confused with each other [48]. To overcome these limitations, phonemes were grouped by their manner of articulation (MOA), which describes the degree to which airflow is obstructed during speech production [49]. Fig. 3 shows the overall spectrum of each phoneme as a function of the electrode index, where each phoneme is grouped into its respective MOA. As the figure demonstrates, phonemes within the same MOA are often spectrally similar to each other, while phonemes from different MOAs have more distinct time-frequency characteristics and are less likely to be confused. Because MOA conveys phonetic information while being less confusable than individual phonemes, we hypothesize that a MOA-based speech enhancement algorithm may still be beneficial.

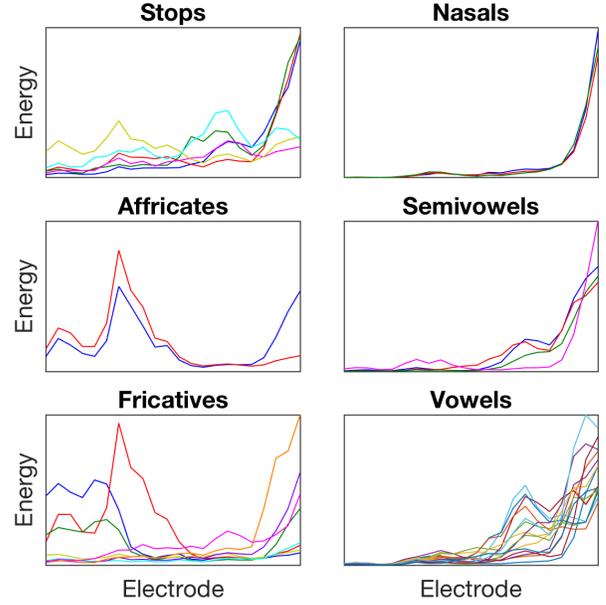

Fig. 3. Energy distributions of each phoneme as a function of electrode index. Each subplot shows the energy of each phoneme within each manner of articulation, where each phoneme is represented with a different line color. Note that frequency decreases as electrode index increases. As the figure illustrates, the phonemes within the same manner of articulation generally have similar energy distributions, although there are some exceptions.

TABLE II
MAPPING PHONEMES TO MANNERS OF ARTICULATION

| Manner of Articulation | Phonemes |
|---|---|
| Stops | /p, t, k, b, d, g/ |
| Affricates | /tʃ, dʒ/ |
| Fricatives | /s, ʃ, f, θ, z, ʒ, v, ð, h/ |
| Nasals | /m, n, ŋ/ |
| Semivowels | /l, ɹ, w, j/ |
| Vowels | /a, æ, ʌ, ɔ, aʊ, aɪ, ɛ, ɚ, eɪ, ɪ, i, oʊ, ɔɪ, ʊ, u/ |
| Non-phoneme | Non-phoneme* |

*Non-phoneme represents instances where silent gaps that were present in the anechoic signal were filled in with reverberant reflections in the corresponding reverberant signal.

The MOAs were defined according to the TIMIT speech corpus, which includes six MOAs: stops, affricates, fricatives, nasals, semivowels, and vowels [50]. An additional class, referred to as non-phoneme, was included to encompass silent gaps as well as instances in the data where silent gaps were filled in with reverberant reflections. The phonemes were mapped to MOA as shown in Table II. The MOA-specific models were trained in a similar way to the phoneme-specific models.

*D. Objective Intelligibility*

The performance of the aforementioned mask estimation algorithms was first evaluated using two objective intelligibility metrics. This analysis was conducted on the previously described testing dataset, which uses HINT sentences [27] convolved with RIRs from an office, stairway, and Aula

Carolina [38]. This analysis compared the objective intelligibility of direct path speech (DP), unenhanced reverberant speech (REV), and enhanced reverberant speech. Enhanced reverberant speech was generated using ratio masks estimated with the 1-layer and 2-layer phoneme-independent models (ERM-1, ERM-2), the MOA-specific models (ERM-MOA), and the phoneme-specific models (ERM-PHN). The estimated ratio masks were compared to the ideal ratio mask (IRM), which provides an upper bound for mask estimation. The ratio masks were compared with the ideal binary mask (IBM), where T-F units are retained only if their SRR exceeds some threshold. The current study defined the IBM using a threshold of -6dB relative to the overall SRR of the reverberant signal, as this threshold has been shown to maximize speech intelligibility in normal hearing listeners [51].

The first objective intelligibility metric is the envelope correlation-based measure (ECM), which has been shown to accurately predict CI users' speech reception thresholds in noisy listening environments [52]. The ECM measures the correlation between the temporal envelopes of the CI processed corrupted signal and clean signal; in this case, the direct path signal. The ECM is calculated by averaging the squared correlation coefficient across all electrode channels, resulting in a number that ranges from 0 to 1. In the current study, the ECM was calculated using unenhanced or enhanced reverberant speech as the corrupted signal and direct path speech as the clean signal. It should be noted that ECM was originally developed for noisy speech, so its correlation with speech intelligibility may not necessarily hold in reverberant conditions. Nevertheless, the ECM is still a valuable tool for assessing the performance of mask estimation models during model development.

The second metric is the speech-to-reverberation modulation energy ratio (SRMR), which was originally developed to predict the intelligibility of reverberant and dereverberated speech in normal hearing listeners [53]. SRMR is defined as the ratio between the average modulation spectral energy in the first four modulation bands (~3-20 Hz) to that in the higher frequency modulation bands [53]. The rationale is that anechoic speech contains low-frequency temporal envelope modulations [54], while reverberant speech contains higher frequency modulations [53]. The current study considers the SRMR-CI, which adapts the SRMR metric specifically for CI users by using a filterbank based on Nucleus CI devices and by only considering modulation frequencies between 4 and 64 Hz, as this range was shown to best predict speech intelligibility in CI users [55]. The SRMR-CI was calculated from the resynthesized time domain signal, which was reconstructed via the overlap-add method using the magnitude of the enhanced signal and phase of the reverberant signal.

### E. Listening Test

Twenty-one subjects aged 18 to 50 (median = 21.0 years) were recruited for the study. This study was approved by the Duke University Institutional Review Board under protocol number 2017-1227 (original protocol approved on September 5, 2002, amendments to conduct remote studies approved on July 21, 2020 and September 25, 2020). All subjects were native speakers of American English and had self-reported normal hearing. Subjects provided informed consent via Research Electronic Data Capture (REDCap) [56], and they were financially compensated for their time. The Nucleus MATLAB Toolbox [57] was used to process speech stimuli based on the ACE processing strategy [43] with 22 electrode channels, and then vocoded with a sine wave vocoder. The vocoded stimuli were equalized to the same root-mean-square (RMS) and were presented bilaterally.

Due to the COVID-19 pandemic, the listening tests were conducted remotely by installing and running a compiled MATLAB script on each subject's personal computer. The experimenter met virtually with the subjects via Zoom to guide them through program installation and the listening test. The subjects were required to share the computer screen for the duration of their study session to ensure that they were on task. Subjects listened to the speech stimuli using their personal headphones, with 14 subjects using earbuds and 7 subjects using over-ear headphones. Each subject was first trained to understand vocoded anechoic speech using sentences from the City University of New York (CUNY) database [58]. Subjects were instructed to type as many words as they could understand, and to type "I don't know" for unintelligible sentences. Additionally, subjects were told that correct spelling mattered. After entering their response, subjects were provided with feedback. During vocoder training, speech stimuli were calibrated using psychophysical means by allowing subjects to adjust the volume to an audible yet comfortable level, which they were instructed to maintain for the remainder of the study. During the testing phase, subjects listened to sentences from the HINT database [27] using the same processing conditions that were used for the objective intelligibility analysis. The order of the testing conditions was randomized, as was the assignment of HINT sentence lists to conditions to minimize order bias.

### F. Analysis

Speech intelligibility was quantified as the percent of correctly identified phonemes across all sentences within a condition. Words in the subject's response that exactly matched words in the target sentence were converted to phonemes and counted toward the total number of correct phonemes. The other words in the subject's response were compared with the words in the target sentence, and a partial match was declared for response-target word pairs with the highest number of matching phonemes. The number of correct phonemes in these partial matches also contributed to the total number of correct phonemes. Sentences where the subject responded "I don't know" were scored as having no correct phonemes. This method of scoring allows for partial credit in instances where subjects correctly identified part of a word but not the entire word.

All statistical tests were conducted using the R statistical software package. The ECMs and SRMR-CI scores were statistically analyzed using a two-way repeated measures ANOVA with within-subjects factors of room and mask, as well as their interaction. The ANOVA model was fit using the R package *rstatix* [59]. Mauchly's test [60] was used to check

TABLE III
OBJECTIVE INTELLIGIBILITY RESULTS

| Metric | Room | REV | ERM-1 | ERM-2 | ERM-MOA | ERM-PHN | IRM | IBM | DP |
|---|---|---|---|---|---|---|---|---|---|
| ECM | Office | 0.679 (0.039) | 0.745 (0.053) | 0.775 (0.042) | 0.755 (0.052) | 0.763 (0.057) | 0.954 (0.019) | 0.865 (0.033) | 1.000 (0.000) |
| | Stairway | 0.638 (0.041) | 0.777 (0.054) | 0.792 (0.045) | 0.786 (0.052) | 0.798 (0.052) | 0.965 (0.016) | 0.895 (0.031) | 1.000 (0.000) |
| | Aula | 0.505 (0.035) | 0.671 (0.060) | 0.669 (0.053) | 0.691 (0.056) | 0.698 (0.058) | 0.965 (0.014) | 0.864 (0.032) | 1.000 (0.000) |
| SRMR-CI | Office | 1.915 (0.546) | 2.294 (0.575) | 2.199 (0.506) | 2.347 (0.568) | 2.449 (0.625) | 3.611 (0.937) | 2.968 (0.830) | 3.871 (0.970) |
| | Stairway | 1.834 (0.519) | 2.667 (0.710) | 2.357 (0.558) | 2.610 (0.642) | 2.716 (0.736) | 3.620 (0.945) | 3.016 (0.813) | 4.002 (0.958) |
| | Aula | 0.923 (0.238) | 1.337 (0.331) | 1.388 (0.333) | 1.379 (0.334) | 1.530 (0.396) | 3.354 (0.879) | 2.493 (0.658) | 3.954 (1.022) |

This table shows the results in terms of objective intelligibility metrics: envelope-based correlation measure (ECM) and speech-to-reverberation modulation energy ratio (SRMR-CI). Results are shown for three reverberant rooms and eight processing conditions: unenhanced reverberation (REV), the phoneme-independent models with 1 or 2 LSTM layers (ERM-1 and ERM-2), the manner of articulation-specific model (ERM-MOA), the phoneme-specific model (ERM-PHN), the ideal ratio mask (IRM), the ideal binary mask (IBM), and the direct path signal (DP).

for sphericity, and the degrees of freedom were adjusted using the Greenhouse-Geisser correction for effects that violated the sphericity assumption. For terms in the ANOVA model that were statistically significant, pairwise comparisons were made using estimated marginal means with Tukey's multiple comparisons correction, using the *emmeans* package [61] in R. The speech intelligibility scores were analyzed in a similar manner, except they were transformed to rationalized arcsine units (RAUs) [62] prior to running inferential statistics. All statistical tests were conducted at a significance level of 0.05.

## III. RESULTS

### A. Objective Intelligibility

Table III compares each mask in the three reverberant rooms using the objective intelligibility metrics. In terms of ECM, the main effects of room [$F(1.9, 462.6) = 1276.1$, $p < 0.0001$, $\eta_g^2 = 0.367$, Greenhouse-Geisser correction] and mask [$F(2.4, 579.2) = 6099.6$, $p < 0.0001$, $\eta_g^2 = 0.897$, Greenhouse-Geisser correction] were statistically significant, as was their interaction [$F(6.3, 1515.0) = 626.8$, $p < 0.0001$, $\eta_g^2 = 0.296$, Greenhouse-Geisser correction]. Overall, all of the estimated ratio masks provided significantly higher ECMs as compared to unenhanced reverberation ($p < 0.0001$). In the office and stairway, ERM-MOA and ERM-PHN generally outperformed ERM-1 but underperformed or provided similar levels of performance to ERM-2. In Aula Carolina, ERM-MOA and ERM-PHN yielded significantly higher ECMs than both ERM-1 and ERM-2. For the ideal T-F masks, the IRM yielded significantly higher ECMs than the IBM in all three rooms ($p < 0.0001$).

In terms of SRMR-CI, the main effects of room [$F(2, 478) = 510.0$, $p < 0.0001$, $\eta_g^2 = 0.207$] and mask [$F(1.6, 375.2) = 1543.5$, $p < 0.0001$, $\eta_g^2 = 0.558$, Greenhouse-Geisser correction] were statistically significant, as was their interaction [$F(6.0, 1437.5) = 339.5$, $p < 0.0001$, $\eta_g^2 = 0.078$, Greenhouse-Geisser correction]. Again, all of the estimated ratio masks provided statistically significant improvements in SRMR-CI compared to unenhanced reverberation ($p < 0.0001$). ERM-PHN provided numerical improvements over all estimated masks in all rooms, and these improvements were statistically significant for all cases except for ERM-MOA in the office and ERM-1 in the stairway. ERM-MOA was less beneficial, only providing statistically significant improvements over ERM-2 in the office and stairway. For the ideal T-F masks, the IRM yielded significantly higher SRMR-CI scores than the IBM in all three rooms ($p < 0.0001$).

### B. Subject Listening Tests

Fig. 4 compares the average speech intelligibility scores on unenhanced reverberant speech and enhanced speech. The main effects of room [$F(2, 40) = 143.1$, $p < 0.0001$, $\eta_g^2 = 0.305$] and mask [$F(3.6, 71.2) = 297.3$, $p < 0.0001$, $\eta_g^2 = 0.769$, Greenhouse-Geisser correction] were statistically significant, as was their interaction [$F(14, 280) = 12.0$, $p < 0.0001$, $\eta_g^2 = 0.210$]. In the office, ERM-2, ERM-MOA, and ERM-PHN did not provide significantly different speech intelligibility scores as compared to REV ($p = 1.000$, $p = 0.989$, and $p = 0.999$, respectively). Although the estimated masks did not provide statistically significant improvements in speech intelligibility, ERM-PHN still benefitted 13 out of 21 subjects by at least 1 percentage point. ERM-1 provided significantly lower performance than REV ($p = 0.035$). In the stairway, ERM-1, ERM-2, and ERM-MOA provided statistically insignificant improvements in speech intelligibility over REV ($p = 0.995$, $p =$ 0.792, and $p = 0.084$, respectively). Despite this, ERM-1 benefitted 12 out of 21 subjects, ERM-2 benefitted 11 out of 21 subjects, while ERM-MOA benefitted 15 out of 21 subjects by at least 1 percentage point. On the other hand, ERM-PHN provided a significant improvement over REV ($p < 0.0001$) as well as significantly outperforming ERM-1, ERM-2, and ERM-MOA ($p < 0.0001$, $p < 0.0001$, $p = 0.009$ respectively). In Aula Carolina, none of the estimated masks provided significantly different speech intelligibility scores as compared to REV ($p = 0.991$ for ERM-1, $p = 1.000$ for ERM-2 and ERM-MOA, $p = 0.136$ for ERM-PHN). Despite this, ERM-PHN still benefitted 14 out of 21 subjects by at least 1 percentage point. In all rooms, the mean speech intelligibility scores for IRM, IBM, and DP were above 90 percent, and were not significantly different from each other ($p > 0.5$ for all pairwise comparisons).

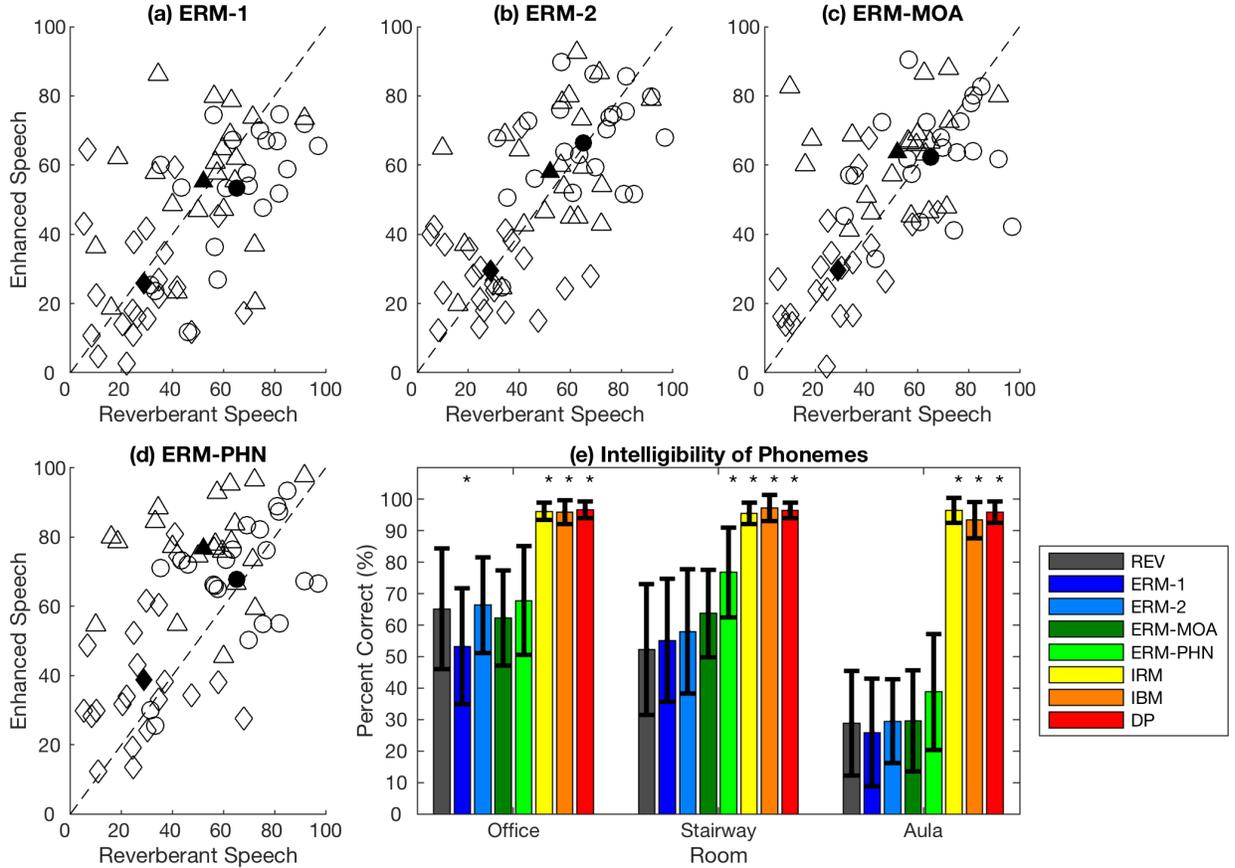

Fig. 4. Subplots (a) through (d) show the relationship between the intelligibility of reverberant speech and enhanced speech for each subject in different masking conditions. Markers above the parity line (dotted black) represent subjects who benefitted from the application of the mask, while markers below the line represent subjects who performed worse with enhanced speech than unenhanced reverberant speech. The open markers represent individual subjects, where the circle, triangle, and diamond denote the office, stairway, and Aula Carolina, respectively. The solid markers denote the mean across all subjects. Subplot (e) shows the percent of correct words across 21 subjects. Results are shown as mean +/- 1 standard deviation. Results are shown for unenhanced reverberation (REV), the phoneme-independent models with 1 or 2 LSTM layers (ERM-1 and ERM-2), the manner of articulation-specific model (ERM-MOA), the phoneme-specific model (ERM-PHN), the ideal ratio mask (IRM), the ideal binary mask (IBM), and the direct path signal (DP). Asterisks denote statistical significance with respect to unenhanced reverberation ($p < 0.05$).

An additional analysis was conducted to determine the relationship between speech intelligibility and objective intelligibility. This analysis was conducted for unenhanced reverberant speech and reverberant speech enhanced using ERM-1, ERM-2, ERM-MOA, ERM-PHN, and IRM in the office, stairway, and Aula Carolina. Fig. 5 shows the average speech intelligibility across all subjects plotted against the objective intelligibility averaged across all sentences in the HINT database. Both the ECM and the SRMR-CI were highly correlated with speech intelligibility. For the ECM, the Pearson correlation coefficient was 0.877, and the Spearman rank correlation was 0.839. The SRMR-CI exhibited higher correlations, with a Pearson correlation coefficient of 0.947, and a Spearman rank correlation of 0.882.

IV. DISCUSSION

The goal of this study was to determine whether ideal phonemic knowledge during T-F mask estimation can benefit normal hearing subjects listening to vocoded speech. Manner of articulation (MOA)-specific and phoneme-specific models were initialized using weights from a trained phoneme-independent mask, and these models were fine-tuned using frames corresponding to the relevant MOA or phoneme. The experimental results demonstrate that the phoneme-specific models provide higher speech intelligibility as compared to the phoneme-independent models and unenhanced reverberant speech. Furthermore, these results demonstrate that a phoneme-based approach to mask estimation results in higher speech intelligibility scores than increasing the depth of a phoneme-independent neural network from one to two LSTM layers. However, these comparisons were statistically significant in only the stairway. The MOA-specific model was less beneficial than the phoneme-specific model, providing insignificantly higher speech intelligibility than the phoneme-independent models and unenhanced reverberant speech in only the stairway. When analyzed on a per-subject basis, the results showed that the phoneme-specific mask benefitted the majority of subjects ($\geq 13$ out of 21) in all three rooms, and the MOA-specific mask benefitted the majority of subjects (15 out of 21) in the stairway. Thus, these statistically insignificant comparisons are likely not due to the lack of consistent improvements in speech intelligibility, but rather due to the large amount of variability between subjects.

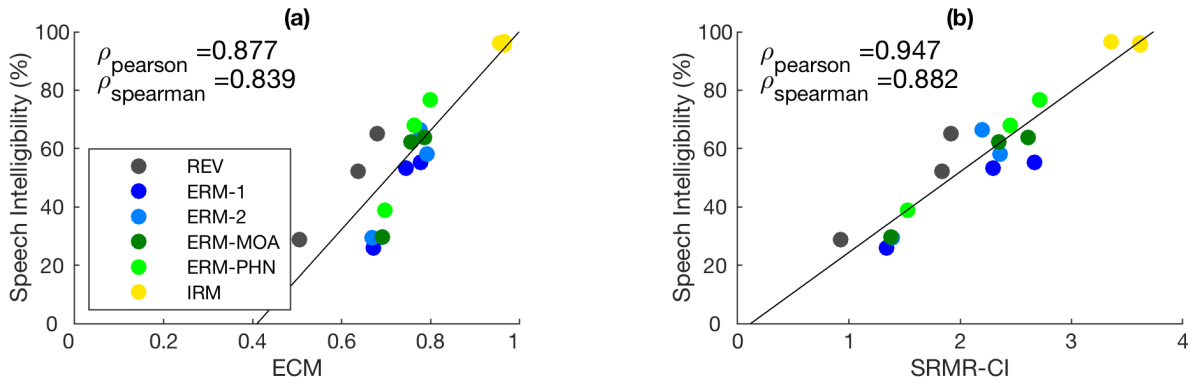

Fig. 5: Relationship between speech intelligibility and objective intelligibility metrics. This analysis was conducted for unenhanced reverberant speech and reverberant speech enhanced using the phoneme-independent mask, MOA-specific mask, phoneme-specific mask, and ideal ratio mask. In each subplot, each marker denotes the average speech intelligibility scores across all subjects in a single test condition plotted against the corresponding objective intelligibility. The black line denotes the best linear fit between speech intelligibility and objective intelligibility. The correlation between speech intelligibility and objective intelligibility is quantified in terms of the Pearson correlation coefficient and the Spearman rank correlation.

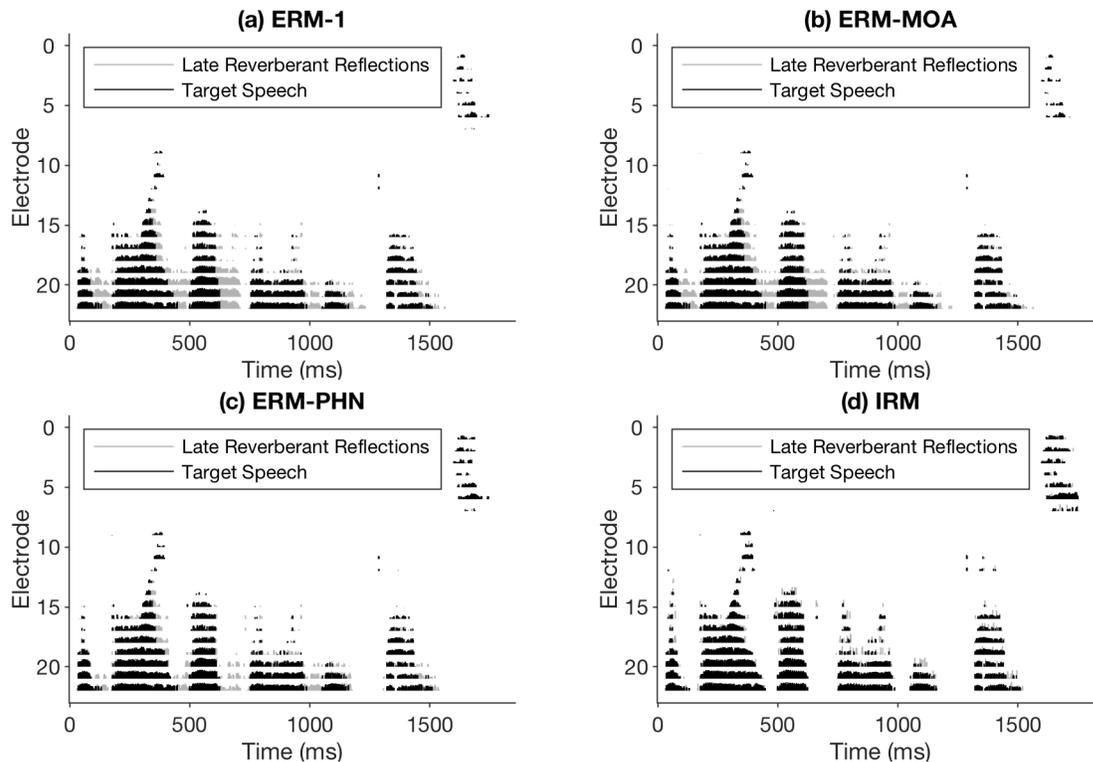

Fig. 6. Electrodograms for the HINT sentence "the boy broke the wooden fence" in a stairway environment generated for reverberant speech enhanced using the conventional estimated reverberant mask (ERM-1), manner of articulation-specific estimated reverberant mask (ERM-MOA), phoneme-specific estimated reverberant mask (ERM-PHN), and the ideal reverberant mask (IRM).

An analysis of objective intelligibility was conducted using the envelope correlation-based measure (ECM), and the speech-to-reverberation modulation energy ratio (SRMR-CI). Overall, both objective intelligibility metrics were highly correlated with speech intelligibility, although there are some notable differences in the trends for objective intelligibility and speech intelligibility. The ECM and SRMR-CI both overestimated the benefit of the phoneme-independent masks, showing higher performance than unenhanced reverberation. However, the listening test results indicated that neither of the phoneme-independent masks provided statistically significant improvements in speech intelligibility in any of the reverberant environments. Additionally, the ECM was lower for the phoneme-specific model than the 2-layer phoneme-independent model in both the office and stairway, with the phoneme-specific model providing a performance benefit only in Aula Carolina. On the other hand, the SRMR-CI showed that the phoneme-specific model provided improvements over both the 1-layer and 2-layer phoneme-independent models in all three reverberant environments, which agrees with the speech intelligibility results. While objective intelligibility metrics can provide insights into predicted speech intelligibility without the need to conduct a study, these metrics do not necessarily translate to speech intelligibility results.

Additionally, a subjective analysis was conducted on a subset of electrodograms in the stairway condition as shown in Fig. 6. This analysis was conducted for the phoneme-independent mask with one LSTM layer, the MOA-specific mask, the phoneme-specific mask, and the ideal ratio mask. As the figure illustrates, the phoneme-independent mask with one LSTM layer primarily removed reverberant reflections in the mid and high frequency electrodes. However, reverberation tended to be retained in the lowest frequency channels, which are the most important channels for speech intelligibility. As compared to the phoneme-independent mask, the phoneme-specific algorithm is better able to recover the silent gaps between phonemes in the low frequency electrodes, especially near 100ms and 500ms. This offers a potential explanation for the superior performance of the phoneme-specific algorithm; however, other factors cannot necessarily be ruled out.

Recall that one of the phoneme classes is *non-phoneme*, which describes instances where reverberant reflections fill in silent gaps that were present in the corresponding anechoic signal. Given that the phoneme-specific algorithm was tested using ideal knowledge of all phoneme classes, including the non-phoneme class, it might appear that the phoneme-specific algorithm is simply restoring the silent gaps. However, the phoneme-specific algorithm still provided larger benefits than the MOA-specific algorithm, which also includes a non-phoneme class. This suggests that the phoneme-specific algorithm is not merely restoring the silent gaps, but rather leveraging the individual differences between phonemes that cannot be captured by the MOA-specific algorithm.

Because the current study was focused on determining the upper bound on the performance using ideal knowledge of the phoneme, it did not prioritize real-time implementation. The MOA-specific and phoneme-specific mask estimation algorithms require classification models to detect the MOA or phoneme. The authors developed a phoneme classification algorithm to categorize phonemes and MOAs using time-frequency features extracted by a CI processor [63]. However, these classification algorithms are far from perfect, especially in reverberant listening environments [63], resulting in the selection of the incorrect MOA-specific or phoneme-specific mask estimation model. Future work is required to investigate the consequences of classification errors on the intelligibility of enhanced speech.

Another limitation is MOA may not necessarily be the best method of grouping phonemes, as phonemes within the same MOA sometimes differ in their frequency content. Apart from MOA, phonemes can also be characterized by their place of articulation (POA), which describes the location in the vocal tract where airflow is obstructed [49]. Phonemes within the same MOA but different POA can have different spectra. For instance, alveolar fricatives /s, z/ and palatal fricatives /ʃ, ʒ/ are characterized by high-intensity noises that occur in high frequency regions, while labiodental fricatives /f, v/ and interdental fricatives /θ, ð/ generally contain lower intensity sounds with flatter spectra [30], [49]. In addition, phonemes can be described as being voiced or voiceless, which describes whether sound production involves vibration of the vocal folds [30], [49]. Voiced phonemes often contain a voice bar, which is a low-intensity periodic signal that occurs in the low frequency region of a spectrogram [30], [49]. MOA represents just one way of grouping phonemes, and other approaches should be investigated.

Finally, it should be noted that while a phoneme-based mask estimation algorithm is beneficial for normal hearing subjects listening to vocoded speech, the results do not necessarily extend to CI users. The sine wave vocoders used in the present study assume that the information presented in each electrode channel is perceived as independent. However, current spread within the cochlea limits the effective number of channels that CI users can perceive, which limits their speech intelligibility in noisy [7] and reverberant environments [64]. Vocoders that simulate current spread via spectral peak smearing result in speech reception thresholds in normal hearing listeners that more closely resemble the results obtained in CI users [65]. In addition, studies in CI users with singled-sided deafness have demonstrated that spectral peak smearing and band-pass filtering produce a sound quality that more closely resembles that of a CI as compared to sine wave vocoders [66]. Aside from current spread, other psychophysical factors not accounted for in the current vocoder simulations include reduced dynamic range [67], quantized current intensity levels [67], frequency-place mismatch [68], and cochlear dead regions [69], which detrimentally impact speech intelligibility in CI users. These aforementioned phenomena can potentially be incorporated in a vocoder simulation. However, given that the performance of a CI is more user-specific and dependent on the individual's etiology of deafness, a study in CI users is ultimately required to understand the full effects of the proposed speech enhancement algorithm.

## V. Conclusion

This study proposed a phoneme-based speech enhancement algorithm for CI users that leverages explicit phonemic knowledge to estimate the time-frequency mask for reverberation suppression. Manner of articulation (MOA) and phoneme-specific mask estimation algorithms were trained separately for each MOA and phoneme, and then tested in a previously unseen reverberant environment assuming ideal knowledge of the MOA or phoneme. The algorithm was tested in normal hearing listeners using vocoded speech, and the results demonstrated that the phoneme-based mask estimation algorithm improved speech intelligibility to a greater extent than a conventional phoneme-independent mask estimation algorithm. The phoneme-based algorithm was tested in the ideal scenario where the phoneme is known perfectly, so the results illustrate the upper bound on performance. A real-time scenario requires a separate model to detect the phoneme, and such a model will make classification errors and select the incorrect phoneme-specific mask, although it is unknown to what extent these classification errors will affect speech intelligibility. Future work will aim to test whether the phoneme-based mask estimation algorithm can still improve speech intelligibility in the non-ideal case where a separate algorithm is used to detect the phoneme.

ACKNOWLEDGMENT

The authors would like to thank the subjects who participated in this study. This research was funded by the National Institutes of Health under grant number R01DC014290-05. The Titan V GPU used in this work was provided by the NVIDIA GPU Grant Program. Support for the Duke Office of Clinical Research to host REDCap was made possible by grant UL1TR001117 from the National Center for Research Resources, a component of the National Institutes of Health (NIH), and NIH Roadmap for Medical Research.